\documentclass[11pt, oneside]{article}   

\usepackage[utf8]{inputenc}
	
\usepackage{geometry}               
\usepackage[parfill]{parskip}   
\usepackage{graphicx}				
\usepackage{palatino}

\usepackage{natbib}
\usepackage{amsmath,amssymb}
\usepackage{url}
\usepackage[hidelinks]{hyperref}

\begin{document}

\textbf{Spatial Bayesian hierarchical modeling of precipitation extremes over a large domain}

Cameron Bracken, Balaji Rajagopalan, Linyin Cheng, Will Kleiber and Subhrendu Gangopadhyay 

\begin{quote}
\textbf{Abstract} 

We propose a Bayesian hierarchical model for spatial extremes on a large domain. In the data layer a Gaussian elliptical copula having generalized extreme value (GEV) marginals is applied. Spatial dependence in the GEV parameters are captured with a latent spatial regression with spatially varying coefficients. Using a composite likelihood approach, we are able to efficiently incorporate a large precipitation dataset, which includes stations with missing data. The model is demonstrated by application to fall precipitation extremes at approximately 2600 stations covering the western United States, $-125$E to $-100$E longitude and $30$N to $50$N latitude. The hierarchical model provides GEV parameters on a $1/8$th degree grid and consequently maps of return levels and associated uncertainty. The model results indicate that return levels vary coherently both spatially and across seasons, providing information about the space-time variations of risk of extreme precipitation in the western US, helpful for infrastructure planning.
\end{quote}

\section{Introduction}

Engineering design of infrastructure such as flood protection, dams, and management of water supply and flood control require robust estimates of return levels and associated errors of precipitation extremes.  Spatial modeling of precipitation extremes not only can capture spatial dependence between stations but also reduce the overall uncertainty in at-site return level estimates by borrowing strength across spatial locations \citep{Cooley:2012ck}. Hierarchical Bayesian modeling of extremes precipitation was first introduced by \citep{Cooley:2012ck} and since has been widely discussed in the literature \citep{Cooley:2010cma,Aryal:2010bw,Atyeo:2012hr,Davison:2012if,Ghosh:2011ec,Reich:2012hc,Sang:2010bt,Sang:2009ii,Apputhurai:2013fv,Dyrrdal:2014bt}. Hierarchical modeling is an alternative to regional frequency analysis providing gridded or pointwise estimates of return levels within a study region \citep{Renard:2011ho}.

Bayesian hierarchical models for spatial extremes have typically been limited to small geographic regions that include on the order 100 stations covering areas on the order of 100,000 km$^2$.  Large geographic regions with many stations present a computational challenge for hierarchical Bayesian models, specifically when computing the likelihood of Gaussian processes (GPs), which for $n$ data points requires solving a linear system of $n$ equations, an $O(n^3)$ operation. Several approaches exist for speeding up GP likelihood computations such as low-rank approximations \citep{Banerjee:2008hc}, composite likelihood methods \citep{Lindsay:1988vt,Heagerty:1998cy,Caragea:2007ib,Varin:2011ug}, spectral methods \citep{Fuentes:2007ew}, restricted likelihoods \citep{Stein:2004fm} and Laplace approximations \citep{Rue:2009fl}. The use of a composite likelihood approach is explored here because we not only wish to estimate covariance parameters but to also produce maps of return levels with small credible intervals. 


Some attempts have been made to model extremes in large regions and with large datasets in a Bayesian hierarchical context. \cite{Reich:2012hc} use a hierarchical max-stable model with climate model output in the east coast to examine spatially varying GEV parameters, with a max-stable process for the data dependence level. \citep{Ghosh:2011ec} model gridded precipitation data over the entire US, for annual maxima at a 5x5 degree resolution (43 grid cells) and copula for data dependence, incorporating spatial dependence directly in a spatial model on the data, not parameters. \citep{Cooley:2010cma} and \citep{Sang:2009ii} model over 1000 grid cells of climate model output using spatial autoregressive models which take advantage of data on a regular lattice to simplify computations. 


The research contributions of this study are as follows.  A Bayesian hierarchical model is proposed which is capable of incorporating thousands of observation locations by utilizing a composite likelihood method. The GEV shape parameter is modeled spatially in order to capture the detailed behavior of extremes in the western US. In addition the model is capable of incorporating stations with missing data with little additional computational overhead. The model is applied to observed precipitation extremes in each season, providing estimated seasonal return levels for the western US. 

In section 2 the general model structure is described. Section 3 describes details of the application to seasonal extreme precipitation in the western US. Results are discussed in Section 4 and Discussion and conclusions are given in Section 5. 




\section{Model structure}

The joint distribution of the $m$ stations in each year is modeled as a realization from a Gaussian elliptical copula with generalized extreme value (GEV) distribution marginals. The copula is characterized by pairwise dependence matrix $\Sigma$. Spatial dependence is further captured through spatial processes on the location $\mu(\mathbf{s})$, scale $\sigma(\mathbf{s})$ and $\xi(\mathbf{s})$ parameters. We assume the parameters can be described through a latent spatial regression where the residual component $w_\gamma(\mathbf{s})$ follows a mean 0, stationary, isotropic Gaussian process (GP) with covariance function $C_\gamma(\mathbf{s},\mathbf{s}^\prime)$ where $\gamma$ represents any GEV parameter ($\mu$, $\sigma$, $\xi$). The corresponding covariance matrix is $C_\gamma(\boldsymbol{\theta}_\gamma)=[C_\gamma(\mathbf{s}_i,\mathbf{s}_j;\boldsymbol{\theta}_\gamma)]_{i,j=1}^m$ where $\boldsymbol{\theta}_\gamma$ represents the covariance parameters. The first layer of the hierarchical model structure is:


\begin{equation}
(Y(\mathbf{s}_1,t),\ldots,Y(\mathbf{s}_m,t))\sim Gcop_m[\Sigma;\{\mu(\mathbf{s}),\sigma(\mathbf{s}),\xi(\mathbf{s})\}]
\end{equation}
\begin{equation}
Y(\mathbf{s},t) \sim \text{GEV}[\mu(\mathbf{s}),\sigma(\mathbf{s}),\xi(\mathbf{s})]
\end{equation}

where $Y(\mathbf{s},t)$ is the response at site $\mathbf{s}$ and time $t$ and $Gcop_m$ stands for ``m-dimensional Gaussian elliptical copula'' with dependence matrix $\Sigma$. The spatial data layer processes in each year are assumed independent and identically distributed. Alternatives to using a copula to construct the joint distribution are an assumption of conditional independence \citep{Cooley:2012ck} and max-stability \citep{Smith:1990va,Schlather:2002eoa,Cooley:2006uj,Shang:2011kr,Padoan:2012it,Sang:a56cidqM}. Marginally, observations are assumed to have a generalized extreme value (GEV) distribution.

The second layer of the hierarchy, also known as the process layer, involves spatial models for the GEV parameters

\begin{equation}
\mu(\mathbf{s}) = \beta_{\mu,0} + \mathbf{x}_\mu^T(\mathbf{s})\boldsymbol{\beta}_\mu(\mathbf{s})+w_\mu(\mathbf{s})
\end{equation}
\begin{equation}
\sigma(\mathbf{s}) = \beta_{\sigma,0} + \mathbf{x}_\sigma^T(\mathbf{s})\boldsymbol{\beta}_\sigma(\mathbf{s})+w_\sigma(\mathbf{s})
\end{equation}
\begin{equation}
\xi(\mathbf{s}) = \beta_{\xi,0} + \mathbf{x}_\xi^T(\mathbf{s})\boldsymbol{\beta}_\xi(\mathbf{s})+ w_\xi(\mathbf{s})
\end{equation}

Where $\beta_{\gamma,0}$ are spatially independent intercept terms, $\mathbf{x}_\gamma^T(\mathbf{s}_i)$ is a vector of $p$ spatially varying predictors and $\boldsymbol{\beta}_\gamma(\mathbf{s})$ is a vector of $p$ spatially varying regression coefficients. 
Covariates will be discussed in Section \ref{sec:covariates}.

The shape parameter $\xi$ is notoriously difficult to estimate, its value determining the support of the GEV distribution. Positive values of $\xi$ indicate a lower bound to the distribution, negative values indicate an upper bound and zero indicates no bounds. In many studies, $\xi$ is modeled as a single value per study area or per region within the study area \citep{Cooley:2012ck,Renard:2011ho,Atyeo:2012hr,Apputhurai:2013fv}. As in \citep{Cooley:2010cma}, we cannot assume that this parameter is constant over the large study region and so it is modeled spatially along with the other GEV parameters. 

For large regions we cannot assume that a constant spatial regression holds for the entire domain and thus must introduce spatial variation in the regression coefficients. The third layer of the hierarchy involves a spatial model for these regression coefficients

\begin{equation}
\boldsymbol{\beta}_\mu(\mathbf{s}) = \sum_{i=1}^k{}c^\mu_i\eta_i^\mu(\mathbf{s})\label{eqn:beta-mu}
\end{equation}
\begin{equation}
\boldsymbol{\beta}_\sigma(\mathbf{s}) = \sum_{i=1}^k{}c_i^\sigma\eta_i^\sigma(\mathbf{s})\label{eqn:beta-sigma}
\end{equation}
\begin{equation}
\boldsymbol{\beta}_\xi(\mathbf{s}) = \sum_{i=1}^k{}c_i^\xi\eta_i^\xi(\mathbf{s})\label{eqn:beta-xi}
\end{equation}

where the $c_i$'s are weights for $k$ basis functions, the $\eta_i$'s, which are distributed throughout the domain. More details are given in section \ref{sec:spatial_reg}.

\subsection{Elliptical copula for data dependence}
Elliptical copulas are a flexible tool for modeling multivariate data \citep{Renard:2011ho,Sang:2010bt,Ghosh:2011ec,Renard:2007fd}. This class of copulas can represent spatial data with any marginal distribution, a particularly attractive feature for extremal data. The Gaussian copula constructs the joint pdf of a random vector ($Y_1,...Y_m$) as 

\begin{equation}
F_{Gaussian}(y_1,\ldots,y_m) = \Phi_\Sigma(u_1,....u_m)
\end{equation}

where $\Phi_\Sigma(u_1,....u_m)$ is the joint cdf of an $m$-dimensional multivariate normal distribution with covariance matrix $\Sigma$, $u_i=\phi^{-1}(F_i[y_i])$,  $\phi$ is the cdf of the standard normal distribution and $F_i$ is the marginal GEV cdf at site $i$. The corresponding joint pdf is 

\begin{equation}
f_{Gaussian}(y_1,\ldots,y_m) = \frac{\displaystyle\prod_{i=1}^mf_i[y_i]}{\displaystyle\prod_{i=1}^m\psi[u_i]}\Psi_\Sigma(u_1,....u_m)
\end{equation}

where $f_i$ is the marginal GEV pdf at site $i$, $\psi$ is the standard normal pdf and $\Phi_\Sigma$ is the joint pdf of an $m$-dimensional multivariate normal distribution.

The dependence between sites is assumed to be a function of distance \citep{Renard:2011ho}. The dependence matrix is constructed with a simple exponential model

\begin{equation}
\Sigma(i,j) = \exp(-||\mathbf{s}_i-\mathbf{s}_j||/a_0)
\end{equation}

where $a_0$ is the copula range parameter. Note that the values in this dependence matrix are not covariances, so by analogy with the variogram, the dependence model is termed the dependogram \citep{Renard:2011ho}. 

\subsection{Spatial regression model}\label{sec:spatial_reg}

For large regions, spatial regression relationships may not hold constant for the entire domain. In this case it is necessary to allow for spatial variation in the spatial regressions for each GEV parameter. Each regression coefficient is represented as a weighted sum of radial basis functions basis functions (Equations \ref{eqn:beta-mu}-\ref{eqn:beta-xi}). The form of these basis functions are

\begin{equation}
\eta_i(\mathbf{s}) = \exp\left(-||\mathbf{s}-\mathbf{s}_i||^2/a_i^2\right)
\end{equation}

where $a_i^2$ is a range parameter determining the spatial extent of the basis function. These basis functions, also known as Gaussian kernels, are placed at points throughout the domain, known as knots, allowing the regression coefficients to vary smoothly in space. 

The knots are placed according to a space-filling design \citep{Johnson:1990du,Nychka:1998gp}. For each GEV parameter, we use 10 knots (Figure \ref{fig:knots}) since based on the author's experience, regression relationships in the western US region tend to hold for regions of a few square degrees. For simplicity, the same knot locations were used for each GEV parameter and the copula but this is not required.

\subsection{Missing Data}

Stations with missing data can be easily incorporated in the model. When the GEV likelihood is computed, years with missing data are simply skipped. With at least 30 years of data at each station, the GEV parameters can be estimated adequately based on only the available data. For simplicity, the copula was fit to only stations with complete data, though missing data could be incorporated by varying the size of the covariance matrix for each year. 

\subsection{Likelihood and priors}

The marginal distribution of $Y(\mathbf{s}_i,t)$ is $\mathrm{GEV}(y(\mathbf{s}_i,t)|\mu(\mathbf{s}_i),\sigma(\mathbf{s}_i),\xi(\mathbf{s}_i))$ where the log-likelihood for some data point $y$ is:

\begin{equation}
\log\mathrm{GEV}(y|\mu,\sigma,\xi)=-\log(\sigma) - (1 + 1/\xi)\log(b) - b^{-1/\xi}  
\end{equation}

where $b = 1 + \xi(y- \mu)/\sigma$. 

Let $\gamma$ represent any of the GEV parameters ($\mu,\sigma,\xi$). The residual Gaussian processes likelihood $p(\mathbf{w}_\gamma | \boldsymbol{\theta}_\gamma)$ is obtained from the multivariate normal density function $\mathbf{w}_\gamma| \boldsymbol{\theta}_\gamma\sim\mathrm{MVN}(\mathbf{0},\Sigma_\gamma)$, where $\Sigma_\gamma=C(\boldsymbol{\theta}_\gamma)$. We use an exponential covariance function with parameters $\delta^2_\gamma$ (the partial sill or marginal variance), $a_\gamma$ (the range) and $\tau^2_\gamma$ (the nugget), so $\boldsymbol{\theta}_\gamma=(\delta^2_\gamma,a_\gamma,\tau^2_\gamma)$. The parametric form of the covariance function is



$$
C(\mathbf{s}_i,\mathbf{s}_j;\boldsymbol{\theta}_\gamma) 
=
\begin{cases} 
      \delta^2_\gamma\exp(-||\mathbf{s}_i-\mathbf{s}_j||/a_\gamma) & i\neq j \\
      \delta_\gamma^2 + \tau_\gamma^2 & i = j
   \end{cases}
$$

We use weakly informative normal priors centered at 0, with a standard deviations as follows: 
0.1 ($\delta^2_\xi, \tau^2_\xi$ ), 
1 ($\delta^2_\mu, \delta^2_\sigma, \tau^2_\mu, \tau^2_\sigma, \beta_0^\xi, c_i^\mu, c_i^\sigma, c_i^\xi; i=1,\ldots,n$), 
10 ($\beta_0^\mu, \beta_0^\sigma$), 
1000 ($a_\mu, a_\sigma, a_\xi, a_0, a_i; i=1,\ldots,n$). For $\xi$ we restrict values to the range $[-0.5,0.5]$, motivated by the typical ranges seen in precipitation data \citep{Cooley:2010cma}.

\section{Estimation}

\subsection{Composite likelihood}

Composite likelihood for spatial data is a method in which the full likelihood is approximated by a set of conditional or marginal likelihoods; see \cite{Varin:2011ug} for a recent review. Conditional approaches construct the composite likelihood as a product of conditional likelihoods for each observation given neighboring observations \citep{Vecchia:1988iv, Stein:2004fm}. Marginal approaches construct the conditional likelihood as a product of joint densities of groups of observations of two or more. The case when a group consists of one observation is known as the independence likelihood, which precludes the computation of spatial dependence parameters \citep{Varin:2011ug}. Composite likelihood methods have also been applied to max-stable spatial processes; see \cite{Sang:a56cidqM} for a recent review.

In our approach, the stations are broken up into $G$ groups each with $n_g$ stations. The marginal composite likelihood estimator ($L_c$) is constructed as a product of the group likelihoods 

\begin{equation}
L_{c}(\boldsymbol\theta|\mathbf{y}_1,\ldots,\mathbf{y}_G) = \prod_{g=1}^GL_g(\boldsymbol\theta|\mathbf{y}_g)
\end{equation}

where $\boldsymbol\theta$ contains covariance parameters and $\mathbf{y}_g$ contains observations from group $g$. This approach is similar to the ``small blocks'' approach from \cite{Caragea:2006wza,Caragea:2007ib}. Approximating the likelihood in this way requires $O(Gn_g^3)$ computations as opposed to $O(n^3)$. An assumption in this approach is that each group is independent, which is expected to introduce some loss of statistical efficiency. As $n_g$ increases (and $G$ decreases) the composite likelihood estimator approaches the true likelihood at the cost of increased computation time \citep{Caragea:2007ib}. The choice of $n_g$ must be a balance between computation time and accuracy. Along these lines, \cite{Caragea:2007ib} suggest that computational efficiency is maximized when $n_g$ is between $m^{1/2}$ and $m^{2/3}$, where $m$ is the total number of stations.  In this application, the composite likelihood approximation is applied to compute the copula likelihood as well as each of the latent GEV parameter residuals. 


\subsection{Composite likelihood group size and distribution}
In order to use a composite likelihood approach we must decide how many stations to use in each group ($n_g$). The number of stations in each group should be small enough so as not to incur substantial computational cost but large enough so that the covariance parameters can be adequately estimated. We used 30 stations per group or approximately 1\% of the total number of stations. The consequences of this choice are explored in Section \ref{sec:group-size-selection}.

We must also choose how stations are to be grouped. Several approaches come to mind such as selecting groups based on climatological regions, elevation bands or a course grid. We chose to group stations randomly, expecting that groups will have a mixture of stations with a range of spatial proximities, allowing for estimation of both small and large scale behavior.

What remains in the model are a few application specific details: selection of the knot locations and the selection of covariates. These are described in the next sections. 

\section{Application to the Western US}

\subsection{Precipitation Data}
\begin{figure}[htbp] 
   \centering
   \includegraphics[width=5in]{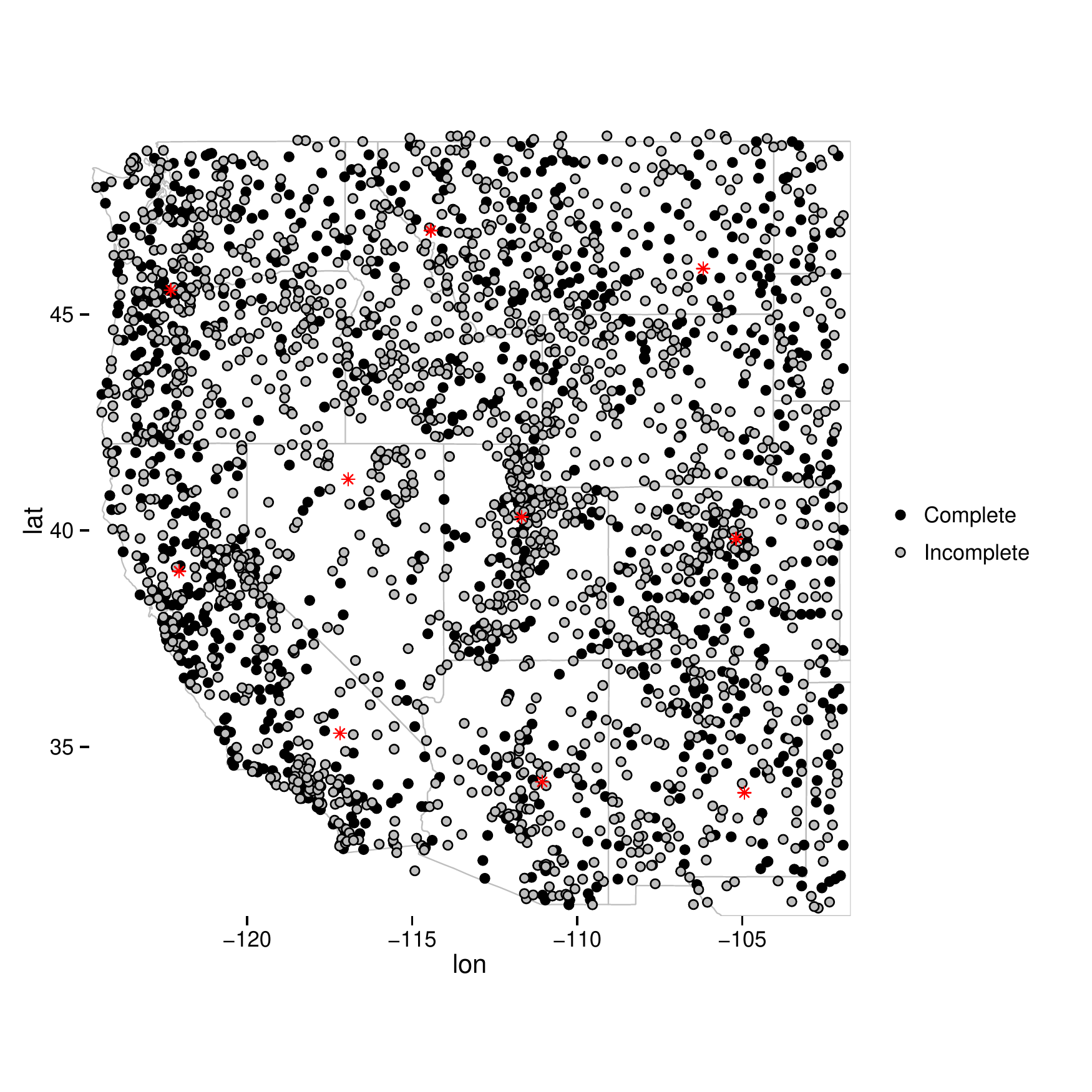} 
   \caption{Station locations with complete data (black solid dots) and station locations with incomplete data (grey filled dots). Red asterisks are knot locations for the spatially varying regression coefficients. }
   \label{fig:knots}
\end{figure}

Daily precipitation data was obtained from the Global Historical Climatology Network (GHCN). We use all available stations in the western US which contain more than 30 years of data from 1950-2013. 3-day maxima were computed fall (SON). For a season to be included for a particular year, we require no more than 25\% of the days be missing. The number of stations included (with the number of complete stations in parentheses) was 2618 (848). Figure \ref{fig:knots} shows the station locations, with solid black points indicating stations with complete data and filled grey points indicating stations with incomplete data. Red asterisks indicate the centers (knots) for the radial basis functions. 

%

\subsection{Covariates}\label{sec:covariates}

For all GEV parameters the same covariates are used, i.e., $\mathbf{x}_\mu(\mathbf{s})=\mathbf{x}_\sigma(\mathbf{s})=\mathbf{x}_\xi(\mathbf{s})=\mathbf{x}(\mathbf{s})$. 
The covariates are elevation and mean seasonal precipitation. Typically, latitude and longitude are used as well but the spatially variation of the regression coefficients precludes this. Covariates were obtained at knot locations, station locations and at a 1/8th degree grid throughout the study area. Elevation data was obtained from the NASA Land Data Assimilation Systems (NLDAS) website \footnote{http://ldas.gsfc.nasa.gov/nldas/NLDASelevation.php} \citep{Xia:2012cg,Xia:2012jn}. Mean seasonal precipitation was computed from the Maurer dataset \citep{Maurer2002}.






\subsection{Implementation}

The model was implemented in the Stan modeling language \citep{stan-software:2015} using the RStan interface \citep{rstan-software:2015}. Stan uses the No-U-Turn Sampler (NUTS), an implementation of Hamiltonian Monte Carlo (HMC) \citep{betancourt:2013,hoffman-gelman:2013}.  The NUTS sampler deals well with highly correlated parameters, tends to need very few warmup iterations and typically produces nearly uncorrelated samples. For these reasons, very long chains are usually not needed, nor is thinning. The tradeoff in using the NUTS sampler in this application was much longer computation time per sample compared to a traditional Metropolis-Hastings or Gibbs sampler.

Three chains of length 3,000 were run, with the first 1000 iterations discarded as warmup, resulting in 6,000 samples for each parameter in each season.  To assess convergence, we compute the $\hat{R}$ statistic to ensure it is below 1.1, as well as visually inspect trace plots.

\subsection{Computation of gridded return levels}

After computing $\boldsymbol{\mu}=[\mu_i]^n_{i=1},\boldsymbol{\sigma}=[\sigma_i]^n_{i=1}$ and $\boldsymbol{\xi}=[\xi_i]^n_{i=1}$ distributions of each GEV parameter are obtained at each 1/8th degree grid cell via conditional simulation. The gridded parameter values are used to compute return levels at each grid cell using the GEV return level formula
$$
z_i(r) = \mu_i + \sigma_i ((-\log (1-1/r))^{-\xi_i} - 1)/\xi_i,
$$

where $r$ is the return period in years (100 years for example). 

\section{Results}

\subsection{Testing the validity of the Gaussian copula}
An implication of the Gaussian copula is that marginal distributions are asymptotically independent, or $P(F_x(X)>p|F_y(Y)>p)\rightarrow0$ as $p\rightarrow1$ \citep{Renard:2007fd}. To test this we conducted asymptotic independence tests 
\citep{Reiss:2007wv}
for all pairs of stations. The null hypothesis of this test is dependence, so setting a significance level of 99\% ensures that stations passing the test exhibit strong asymtotic dependence. At the 99\% significance level, 0.15\% of pairwise stations exhibited dependence, less that the 1\% expected from chance. In addition we examined plots of the station locations when dependence was indicted by the test. These plots did not show any discernible spatial pattern of dependence, for example dependent stations did not tend to fall near each other.

\subsection{Group size selection}\label{sec:group-size-selection}
To demonstrate that the selection of group size has little effect on return levels, a small experiment is conducted. We run the model for a region encompassing most of the state of Oregon, using 4 knots. The group size is set to be 2, 5, 10, 15, 20 and 30 stations representing approximately 1\%, 2\%, 4\%, 6\%, 8\% and 13\% of the total number of stations respectively. The same 240 stations (60 complete, 180 incomplete) are used in each model run. 

\begin{figure}[htbp] 
   \centering
   \includegraphics[width=\textwidth]{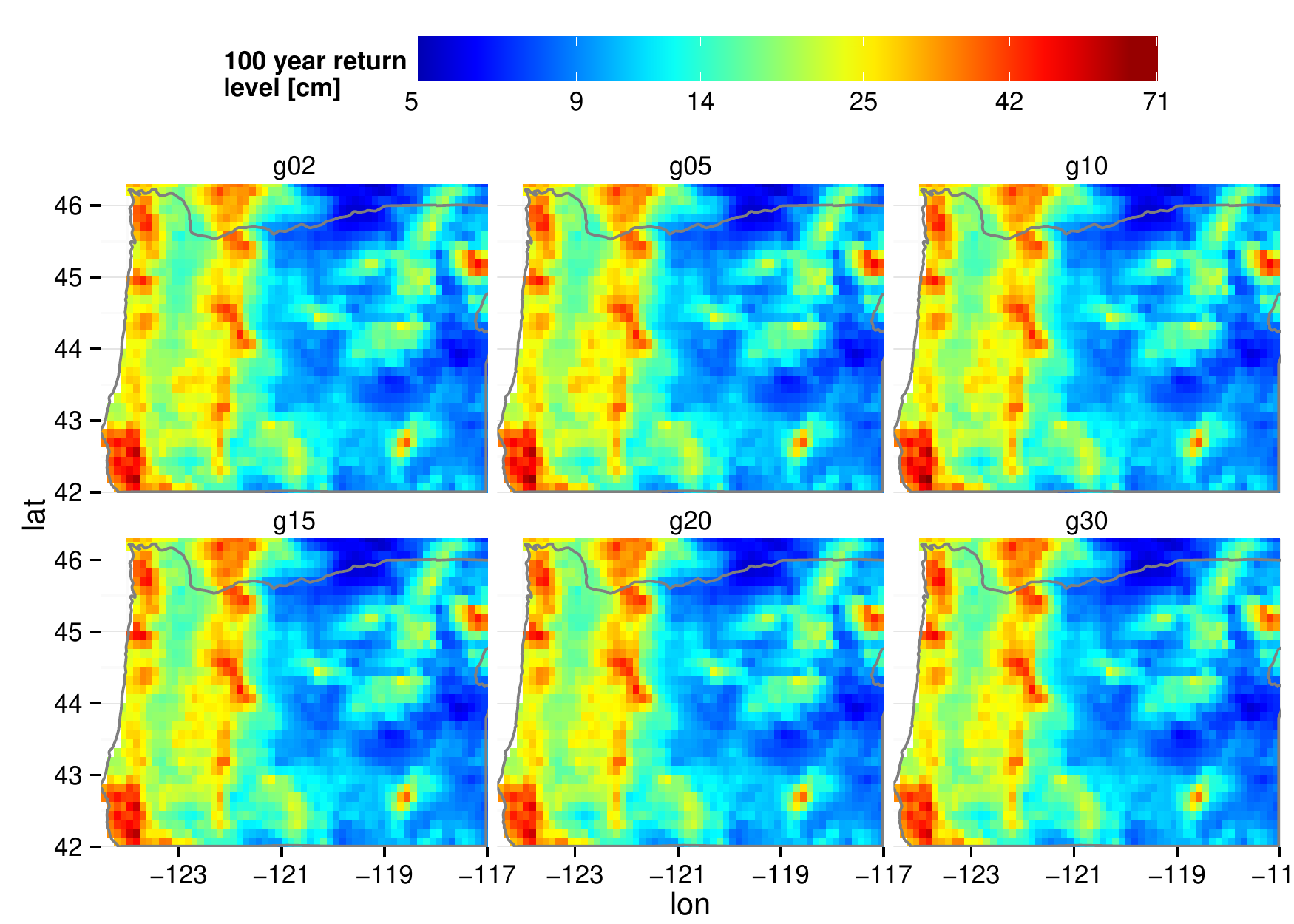}
   \caption{Median return levels using a group sizes of 2, 5, 10, 15, 20 and 30. Note the logarithmic color scale.}
   \label{fig:rl-group-size}
\end{figure}

Figure \ref{fig:rl-group-size} shows the median return level for each model run. The results are nearly identical for this range of group sizes, indicating that median return levels are not sensitive to the choice of group size. Credible intervals of return levels (not shown) were quite similar as well, with credible intervals decreasing as group size is increased indicating that a larger group size yields more accurate results, as expected. In light of this we chose a group size of 30 for the large domain which provides both a diversity in the distribution of stations within a group but is small enough to not significantly hider computation. 

\subsection{Gridded return levels}
\begin{figure}[!h] 
   \centering
   \includegraphics[width=\textwidth]{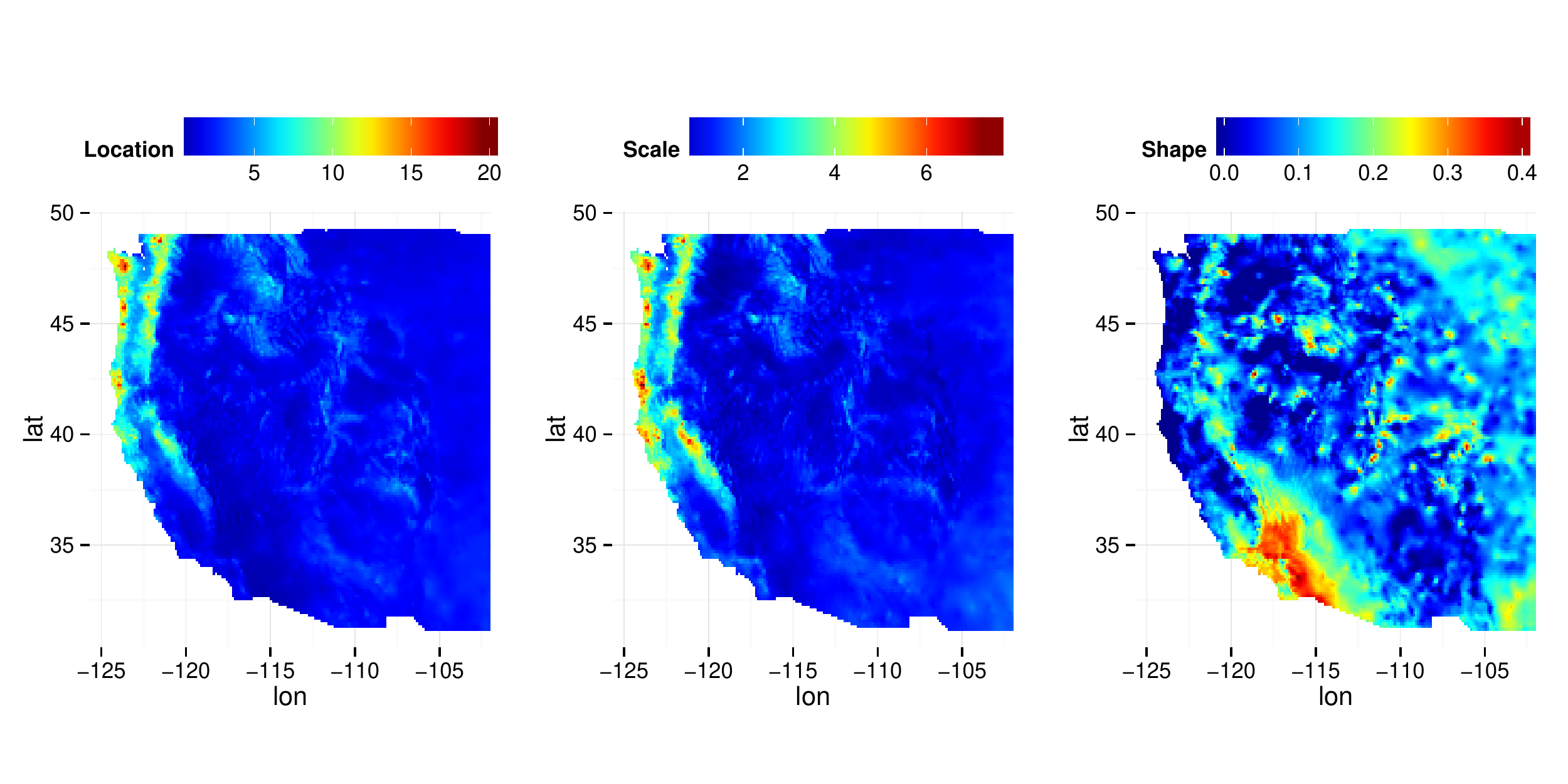}
   \caption{Median of underlying GEV parameters, location ($\mu$), scale ($\sigma$) and shape ($\xi$).}
   \label{fig:shape}
\end{figure}

Figure \ref{fig:shape} shows the median of the GEV parameters after interpolation by conditional simulation. The location and shape fields are highly correlated; locations with higher average extreme precipitation tend to have more variability in these extremes. Values of $\xi$ are always positive, indicating a heavy upper tail. The southern coastal region in California in the summer indicates a very heavy upper tail.  

\begin{figure}[htbp] 
   \centering
   \includegraphics[width=\textwidth]{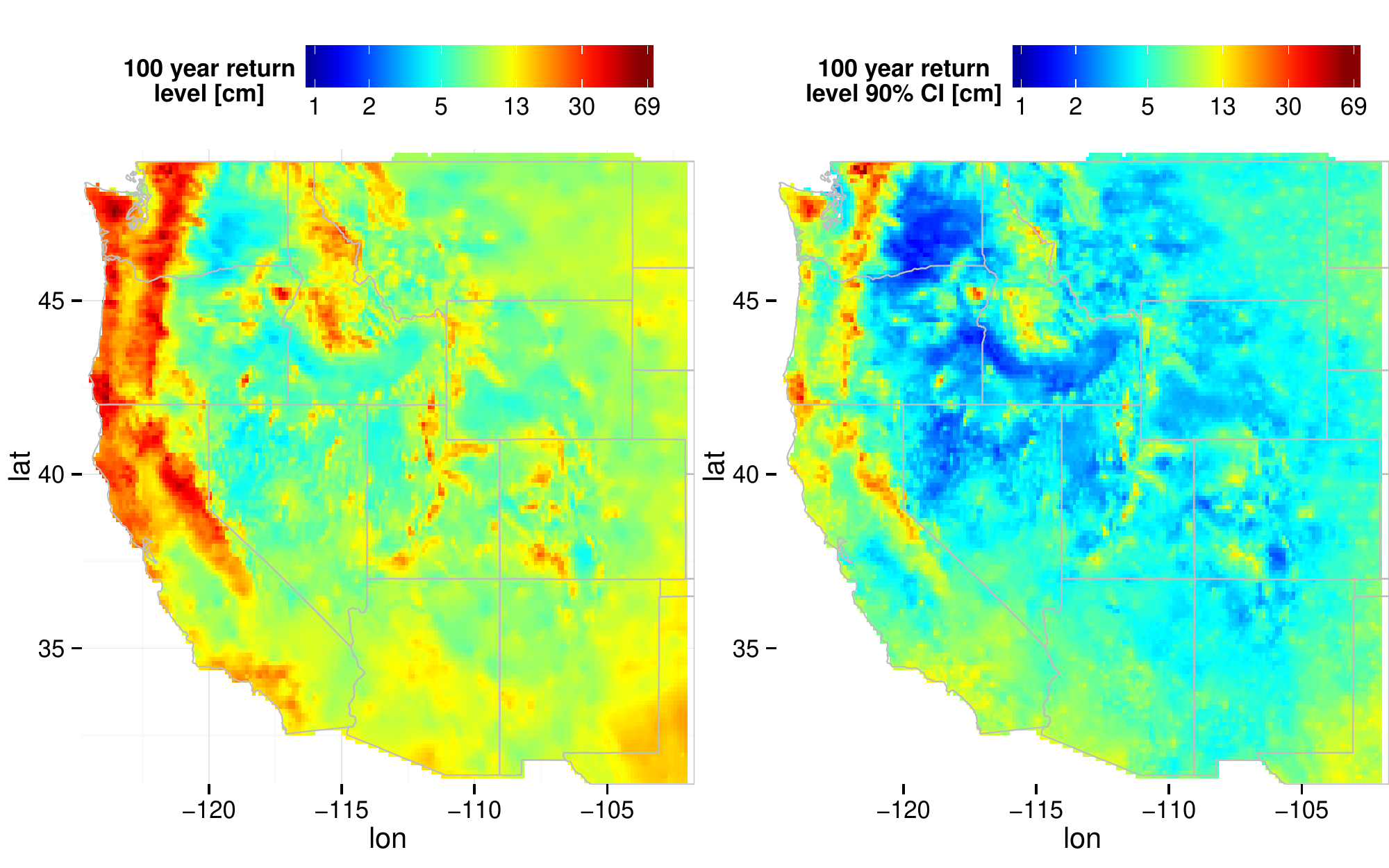}
   \caption{Median 100-year return levels for fall (left) and width of corresponding 95\% credible interval (right). Note the logarithmic color scale.}
   \label{fig:rl-fall}
\end{figure}





\subsection{Validation}

\begin{figure}[!h] 
   \centering
\includegraphics[width=.5\textwidth]{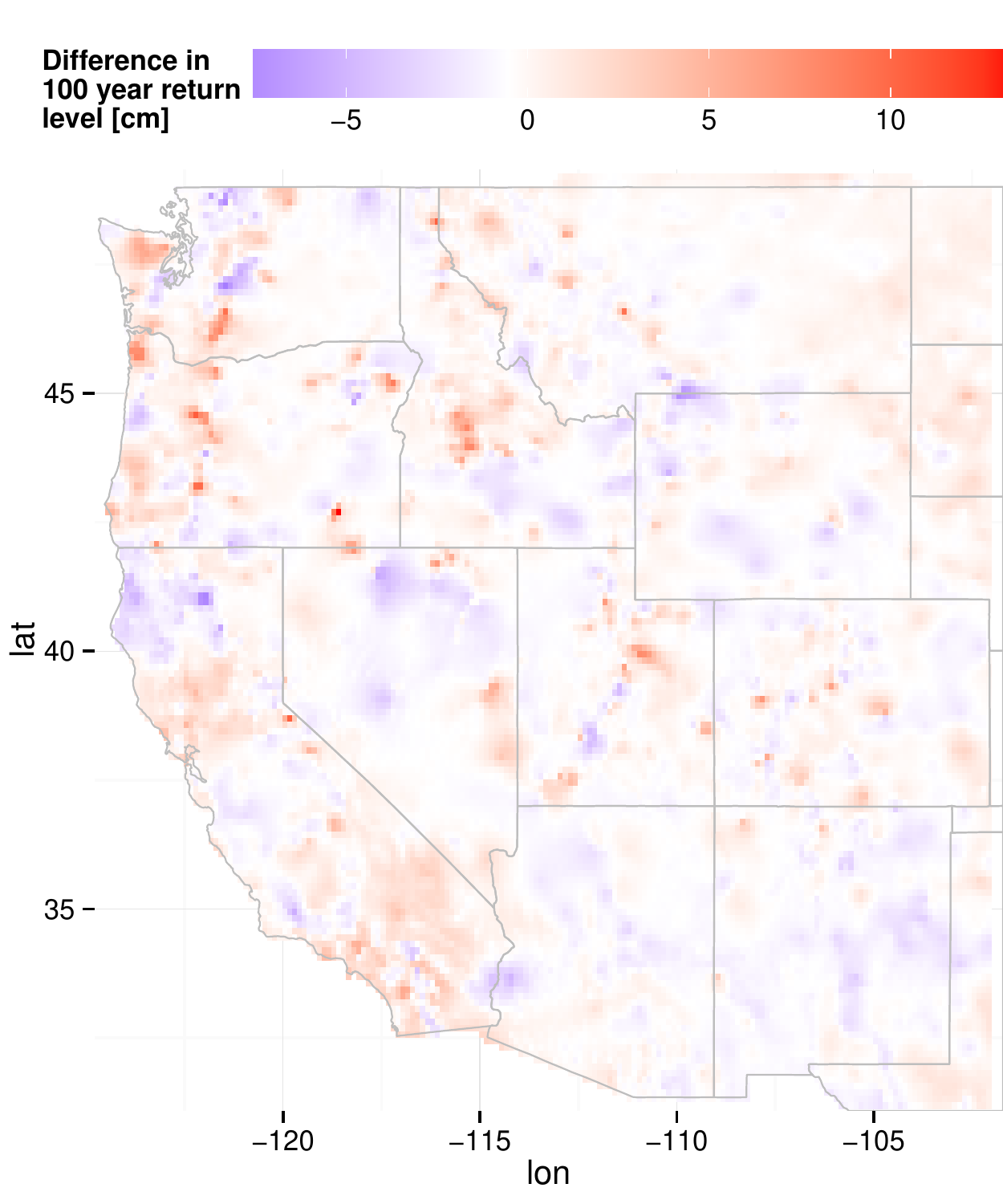}
  \caption{Difference between 50th percentile return levels from the full model and the validation model dropping 35\% of the data.}
   \label{fig:rl-diff}
\end{figure}

Cross validation was conducted by dropping 885 stations or approximately 35\% of the total stations. Gridded return levels were computed for this subset of data.  Figure \ref{fig:rl-diff} shows the difference between the median return level for the full data and subset data. The differnce map shows some spatial coherence but none that indicates any strong bias in a single region (states for example). The largest differences occur in areas in the northwest where influential stations were dropped randomly. 

\subsection{A case for composite likelihood}\label{sec:gpp}

\begin{figure}[!h] 
   \centering
   \includegraphics[width=\textwidth]{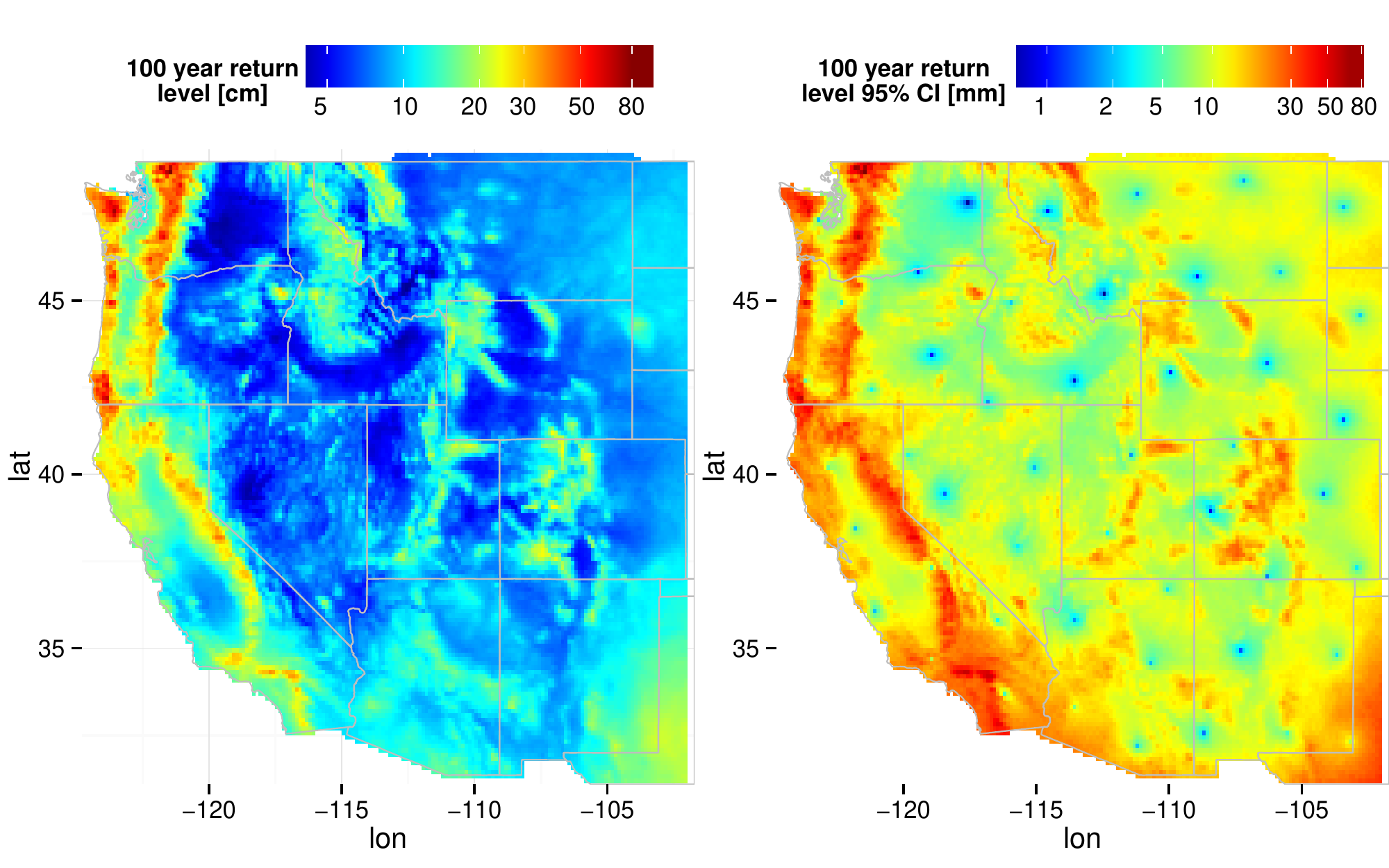}
   \caption{Return levels maps produced using latent gaussian predictive processes.}
   \label{fig:gpp}
\end{figure}

To highlight the usefulness of the composite likelihood approach for this application we present results using a Gaussian predictive process (GPP) model \citep{Banerjee:2008hc} for the latent GEV parameter processes (Figure \ref{fig:gpp}). A Gaussian predictive process model approximates the likelihood at a small set of knots to reduce the dimension of the covariance matrix and the computational burden of inverting it. We originally set out using GPPs for this application but switched to a composite likelihood approach when we realized the uncertainty was unacceptably large away from knot locations. 

The median return levels with the GPP approach were nearly identical to those from the composite likelihood method (Figure \ref{fig:rl-fall}) but large differences are apparent when looking at the credible intervals of the return levels. Clear artifacts are present at the locations of knots, where uncertainty is greatly reduced. Uncertainty away from knot locations 
is typically large, rendering this method much less useful than the composite likelihood approach. 

\section{Discussion and conclusions}

We describe a general Bayesian hierarchical model for extreme data observed over space and time. The data is assumed to originate from a Gaussian elliptical copula having generalized extreme value (GEV) marginal distributions. Spatial dependence is further captured by Gaussian processes on the three GEV parameters (location, scale and shape). Using a composite likelihood approach, we are able to incorporate 2595 observation locations with 54 years of data. With spatially varying regression coefficients, the model can be applied to arbitrarily large regions. The model was applied to extreme 3-day precipitation in fall in the western United States, a climatologically and geographically diverse region. The model was fit using a standard Bayesian methodology, implicitly capturing uncertainty in the parameter estimates and spatial predictions.


In Section \ref{sec:gpp} we briefly examine results for the same region using a Gaussian predictive process (GPP) model for the latent GEV parameters. In this application, GPPs produced unreasonably large posterior credible intervals when moving away from knot locations. In light of this we recommend a composite likelihood approach for regions of equal or larger size than the western US. 

A crux of this model is the use of appropriate spatial covariates. Mean seasonal precipitation (MSP) had a correlation of 95\% with the MLE estimates of $\mu$ and 75\% with the MLE estimates of $\sigma$. Even with spatially varying regression coefficients, appropriate covariates are key. The covariates here helped in generating realistic spatial variability and helped to reveal a complex spatial pattern for the shape parameter, $\xi$. The strongest covariate for $\xi$ was elevation. The spatial variability in $\xi$ shows that it is inappropriate to model without spatial variation for anything but the smallest regions.  

A number of extensions can be made to this framework. The most obvious extension is to allow temporal variation in the GEV parameters by including temporal covariates. While this extension remains infeasible for the size of the current study region, it may be feasible for smaller regions, say a single state or moderate sized river basin. Additional spatial covariates could be included; for example, seasonal temperature, winds or evapo-transpiration. A model such as the one presented here can be used to investigate changes in risk under specific climate regimes (i.e. ENSO); one would simply include the mean seasonal precipitation field from strong El Ni\~no or La Ni\~na years. Because we incorporate a data layer, this model could be used to simulate realistic fields of extremes under specific climate regimes.  Finally, we plan to explore the linking of streamflow data into the hierarchy, so that streamflow extremes can be simultaneously estimated.

\section{Acknowledgments}
Funding for this research by a Science and Technology grant from Bureau of Reclamation is gratefully acknowledged. Kleiber’s portion was supported by NSF DMS-1406536. This work utilized the Janus supercomputer, which is supported by the National Science Foundation (award number CNS-0821794) and the University of Colorado Boulder. The Janus supercomputer is a joint effort of the University of Colorado Boulder, the University of Colorado Denver and the National Center for Atmospheric Research. The authors are thankful for support from the Janus supercomputer staff at the University of Colorado. 

Pre- and postprocesseing analysis was conducted using the R language \citep{Team:wf}.

Data is available at: \url{http://bechtel.colorado.edu/~bracken/spatial_extremes/}.

\bibliography{references}
\end{document}